# g-factor calculations from the generalized seniority approach

*Bhoomika* Maheshwari[1,*], and *Ashok Kumar* Jain[2]

[1]Department of Physics, Banasthali Vidyapith, 304022 Banasthali, India
[2]Department of Physics, Indian Institute of Technology Roorkee, 247667 Roorkee, India

**Abstract.** The generalized seniority approach proposed by us to understand the B(E1)/B(E2)/B(E3) properties of semi-magic nuclei has been widely successful in the explanation of the same and has led to an expansion in the scope of seniority isomers. In the present paper, we apply the generalized seniority scheme to understand the behavior of g-factors in semi-magic nuclei. We find that the magnetic moment and the g-factors do show a particle number independent behavior as expected and the understanding is consistent with the explanation of transition probabilities.

## 1 Introduction

Recent experimental advances towards the neutron drip line provide us opportunities to test the known theoretical ideas and understand the systematics and trends in newer contexts. We have recently developed a simple generalized seniority scheme for multi-j configurations. A new kind of seniority isomers and their related seniority selection rules have been established for the first time[1]. This generalized seniority scheme is quite successful in explaining the reduced transition probabilities and the corresponding half-lives of the semi-magic isomers in different mass regions[1-5].

Sn isotopes present the longest semi-magic isotopic chain from the doubly magic $^{100}$Sn to the next doubly magic $^{132}$Sn and beyond. The magnetic dipole moments (g-factors) of a nuclear state depend on the orbital and spin angular momentum contributions of protons and neutrons involved. So the configuration mixing and the single particle structure near the Fermi surface affect the g-factor significantly. In the present work, we extend the generalized seniority calculations to obtain the g-factors, particularly in the neutron-rich Sn isotopes.

Several theoretical attempts have predicted the g-factors (magnetic moments) of the first excited $2^+$ states in Sn isotopes [6-9]. However, the experimental information still remains incomplete towards $^{100}$Sn and $^{132}$Sn, the two extreme doubly magic nuclei. The g-factor of the first excited $2^+$ states in the stable and even-even Sn isotopes were measured by Hass et al [10]; they reported a positive and almost zero value of the g-factor in the case of $^{118}$Sn and negative values for the neighboring nuclei $^{116,120}$Sn on either side. In this paper, we present the g-factor calculations for the first excited $2^+$ states in Sn isotopes by using our generalized seniority approach.

Besides this, we also discuss the g-factor calculations for the $11/2^-$ states and the $10^+$ high spin isomers in Sn isotopes. The calculated results are in line with the experimental data. Predictions for the $27/2^-$ isomers have also been made.

## 2 Theoretical Framework

The magnetic moment of identical nucleons in pure-$jj^n$ configuration is given by

$$\mu = \sum_{i=1}^{n} g j_i = g \sum_{i=1}^{n} j_i = gJ \qquad (1)$$

It is well known that the magnetic moments and g-factors show a particle number independent behavior for identical nucleons in single-j shell of a pure-seniority scheme [11]. Hence, g-factors of all the states arising from a pure $j^n$ configuration of identical nucleons should be equal to the g-factor of a single-j nucleon in the $j^n$ configuration.

As per our previous extension of single-j seniority scheme to the multi-j generalized seniority scheme [1], we can write the magnetic moment of identical nucleons in the mixed configuration $\tilde{j}^n$ as follows

$$\mu = g \sum_{i=1}^{n} \tilde{j}_i = gJ \qquad (2)$$

where $\tilde{j} = j_1 \otimes j_2 \otimes ...$ is the multi-j, mixed configuration. This means that the g-factors of a mixed configuration also show a particle number independent behavior in the generalized seniority scheme. Hence, g-factors of all the states arising from a given mixed configuration having identical nucleons must be equal to the g-factor of a single nucleon arising from the same mixed configuration.

## 3 Calculations and results

We present in Fig. 1, the experimental and the calculated g-factor trends for the first excited $2^+$ states in Sn isotopes. We fit one of the experimental data of magnetic moments and calculate the g-factor using the formula in Equation (2) for the full chain by using the two mixed configurations $\tilde{j} = g_{7/2} \otimes d_{5/2} \otimes s_{1/2} \otimes d_{3/2}$, and $\tilde{j} = d_{5/2} \otimes s_{1/2} \otimes d_{3/2} \otimes h_{11/2}$, respectively (before and after the middle of the shell). The calculated g-factor results and configurations are consistent with our previous results on two B(E2) parabolas[2]. All the experimental data have been taken from Stone's table [12]. Two parabolas in B(E2) curve highlight the difference in the configuration mixing, before and after the middle for the generation of the $2^+$ states in Sn isotopes [2]. Therefore, g-factors are found to be of different signs before and after the middle. The positive g-factor can be attributed to the g7/2 before the middle, whereas the negative value can be understood as the dominance of h11/2 orbital after the middle.

---

[*]Corresponding author: bhoomika.physics@gmail.com

We then present in Fig. 2 and 3, the g-factor calculations for the $11/2^-$ states and $10^+$ isomers in Sn isotopes, respectively. Note that the $11/2^-$ states correspond to the unique-parity h11/2 orbital of the 50-82 neutron valence space. On the other hand, we use the configuration mixing $\tilde{j} = s_{1/2} \otimes d_{3/2} \otimes h_{11/2}$, as reported in our previous papers [1, 3], for the g-factor calculations of $10^+$ isomers in Sn isotopes. We get the expected trend (particle number independent), which matches with the experimental data well.

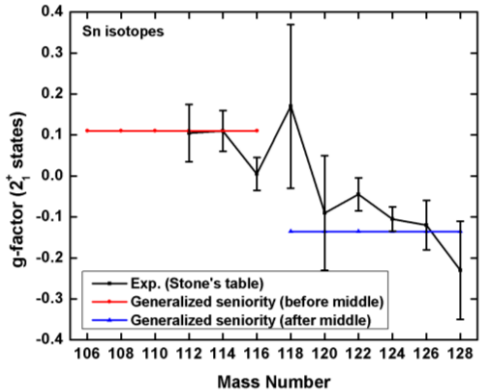

**Fig. 1.** Experimental [12] and calculated g-factor trend for the first excited $2^+$ states in Sn isotopes.

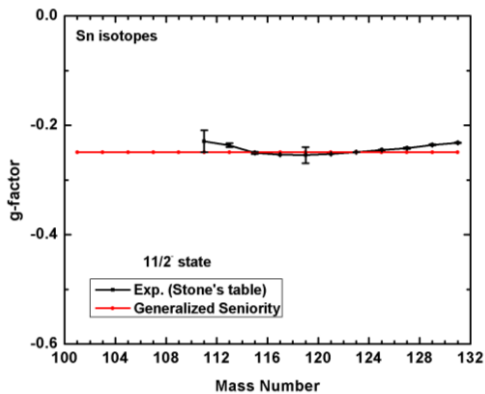

**Fig. 2.** Experimental [12] and calculated g-factor trend for the $11/2^-$ states in Sn isotopes.

However, the data on $27/2^-$ isomers is not known till date. We expect the g-factor of this state to be of the same order as the $10^+$ isomers, since both follow the same configuration mixing [8].

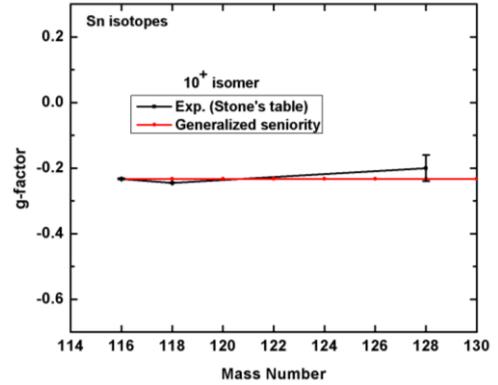

**Fig. 3.** Experimental [12] and calculated g-factor trend for the $10^+$ isomers in Sn isotopes.

## 4 Conclusion

In the present work, we present the g-factor calculations for the first excited $2^+$ states, $11/2^-$ states and $10^+$ isomers in Sn isotopes by extending the usage of our scheme. We find that the states show a particle number independent trend, as per the seniority and generalized seniority scheme. The calculated results explain the experimental data quite well. Prediction for the g-factors of the 27/2- isomers in Sn isotopes has been made.

## References

1. B. Maheshwari, and A. K. Jain, Phys. Lett. B **753**, 122 (2016)
2. B. Maheshwari, A. K. Jain and B. Singh, Nucl. Phys. A **952**, 62 (2016)
3. A. K. Jain, and B. Maheshwari, Nuclear Phys. Review **34**, 73 (2017)
4. A. K. Jain, and B. Maheshwari, Physica Scripta **92**, 074004 (2017)
5. B. Maheshwari, S. Garg, and A. K. Jain, Pramana-Journal of Physics **89**, 75 (2017)
6. J. Terasaki, J. Engel, W. Nazarewicz, and M. Stoitsov, Phys. Rev. C **66**, 054313 (2002)
7. B. A. Brown et al., Phys. Rev. C **71**, 044317 (2005)
8. A. Ansari, and P. Ring, Phys. Lett. B **649**, 128 (2007)
9. L. Y. Jia, H. Zhang, and Y. M. Zhao, Phys. Rev. C **75**, 034307 (2007)
10. M. Hass, C. Broude, Y. Niv, A. Zemel, Phys. Rev. C **22**, 97 (1980)
11. I. Talmi, Simple Models of Complex Nuclei, Harwood Academic (1993)
12. N.J. Stone, www-nds.iaea.org/publications, INDC(NDS)-0658, Feb. (2014)